\documentclass[%
,preprint%
,showpacs
 ,secnumarabic%
,amssymb, amsmath,nobibnotes, aps, prl
]{revtex4}
\usepackage{docs}%
\usepackage{bm}%
\usepackage{graphicx}
\expandafter\ifx\csname package@font\endcsname\relax\else
 \expandafter\expandafter	
 \expandafter\usepackage
 \expandafter\expandafter
 \expandafter{\csname package@font\endcsname}%
\fi

\def\e{{\rm e}}
\def\vector#1{{\boldsymbol{#1}}}

\def\vp{{\vector p}}
\def\vq{{\vector q}}

\def\vR{{\vector R}}

\def\vv{{\vector v}}

\def\Tc{{T_c}}

\def\He3{\mbox{${\rm ^3He}$}}
\def\PrOsSb{\mbox{${\rm PrOs_4Sb_{12}}$}}
\def\ReZr{\mbox{${\rm Re_6Zr}$}}
\def\URu{\mbox{${\rm URu_2Si_2}$}}
\def\LaNi{\mbox{${\rm LaNiC_2}$}}

\def\eq.#1{Eq.~(\ref{#1})}
\def\eqs.#1{Eqs.~(\ref{#1})}

\def\Hc2{{H_{\rm c2}}}

\def\difHc2{{H'_{\rm c2}}}
\def\dif2Hc2{{H''_{\rm c2}}}

\newcommand\Equation[2]{
\begin{equation}\label{#1} 
#2
\end{equation}
}

\begin{document}

\title{Fermi surface effect on spontaneous breaking of time-reversal
symmetry in unconventional superconducting films}

\author{Nobumi Miyawaki and Seiji Higashitani}

\affiliation{Graduate School of Integrated Arts and Sciences, Hiroshima
University, Kagamiyama 1-7-1, Higashi-Hiroshima 739-8521, Japan}

\date{\today}

\begin{abstract}
 We propose a mechanism that helps stabilize a superconducting state with
 broken time-reversal symmetry, which was predicted to realize in a $d$-wave
 superconducting film [A. B. Vorontsov, Phys. Rev. Lett. \textbf{102},
 177001 (2009)]. In this superconducting phase, the time-reversal symmetry
 breaking is accompanied by spontaneous breaking of the translation symmetry
 along the film surface. We examine how the normal-superconducting phase
 boundary in the thickness-temperature phase diagram of the film is modified
 depending on the Fermi surface shape. In particular, the nonuniform
 superconducting phase is found to substantially extend to a smaller thickness region
 in the phase diagram when the Fermi surface satisfies a nesting
 condition. We demonstrate this Fermi surface effect using a
 square-lattice tight-binding model.
\end{abstract}


\maketitle

Spontaneous time-reversal (TR) symmetry breaking in superconductors and
superfluids was first established for the $p +ip$ pairing state in the A phase
of superfluid $^3$He \cite{Vol90}. An analogous chiral $p$-wave state has
been discussed as a promising candidate for quasi-two-dimensional (Q2D)
superconductor Sr$_2$RuO$_4$ \cite{Mae12}. The possibility of TR symmetry
breaking is also discussed for heavy fermion superconductors
$\PrOsSb$~\cite{Aok03} and $\URu$~\cite{Kas07} and for noncentrosymmetric
superconductors $\LaNi$~\cite{Hil09} and $\ReZr$~\cite{Sin14}.

Recent theoretical studies of the surface effects in unconventional pairing
states have aroused renewed interest in TR symmetry-breaking states.
Interestingly, the surface in superfluids and
superconductors provides a mechanism responsible for spontaneous symmetry
breaking. For example, when the B phase of superfluid $^3$He is confined in
the film geometry, pair breaking at the film surface gives rise to spontaneous
breaking of the translation symmetry along the surface \cite{Vor07}. Vorontsov
found that a $d$-wave superconducting (SC) film with pair-breaking surfaces
can exhibit superconductivity that breaks not only the translation
symmetry but also the TR symmetry \cite{Vor09}. Hachiya et al.
examined the stability of this TR symmetry-breaking phase against external
magnetic fields \cite{Hac13}. Very recently, a TR symmetry-breaking state
accompanied by an unusual vortex pattern was predicted for a small $d$-wave
SC grain \cite{Hak14}.

In this Rapid Communication, we address the effect of the Fermi surface shape on
the phase transition of SC films. This study is motivated by previous
theoretical studies \cite{Shi94,Shi99,Miy14} on the Fulde-Ferrell-Larkin-Ovchinnikov
(FFLO) SC state that is stabilized in bulk materials under a strong applied magnetic
field.  The gap function in the FFLO state is reminiscent of those in
nonuniform states proposed for the films of superfluid $^3$He \cite{Vor07}
and the $d$-wave superconductor \cite{Vor09}.
One of the remarkable characteristics of the FFLO state is that
its upper critical field $H_{c2}$ strongly depends on the Fermi surface shape \cite{Shi94,Shi99,Miy14}.
The question
then naturally arises how the Fermi surface
shape affects the phase transition in restricted geometries, where the
instability is triggered not by an external field but by surface-pair breaking.
We examine this problem in the context of the phase transition from the
normal (N) state to the TR symmetry-breaking SC state predicted for a
$d$-wave SC film \cite{Vor09}.

\begin{figure}[h]
\includegraphics[width=8cm]{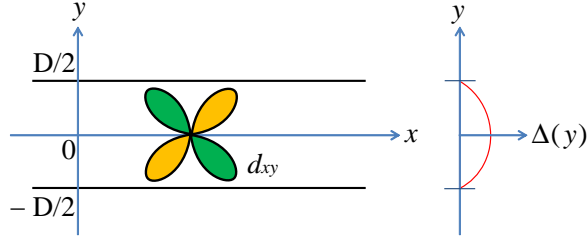} \caption{\label{fig:Dfilm} (Color online) A
$d$-wave SC film with a distorted gap function. The gap function is strongly
suppressed at the surface because of the destructive interference effect caused
by quasiparticle scattering at the film surfaces.}
\end{figure}

The SC film considered here is the same as that in Ref.\ \cite{Vor09} (Fig.\
\ref{fig:Dfilm}). Two parallel surfaces are located at $y = \pm D/2$ and are
assumed to be specular. The $x$-axis is taken along the surface, and the
$z$-direction is assumed to be uniform. We are interested in the phase boundary
at which the N state becomes unstable against a Q2D SC state with the
$d$-wave gap function \cite{Vor09}
\begin{align}
 \Delta(\bm{R}, \bm{p}) = \gamma(\bm{p})\Delta(\bm{R})
\end{align}
with
\begin{align}
 &\gamma(\bm{p}) = \sqrt{2} \sin2 \varphi_{\vp},\\
 &\Delta(\bm{R}) = \Delta(y)e^{iq_xx}.
\end{align}
Here, $\vR =(x,y)$ denotes the spatial coordinate, and $\varphi_{\bm{p}}$ is
the azimuthal angle specifying the direction of momentum $\bm{p}$. The basis
function $\gamma(\bm{p})$ is normalized as
\begin{align}
 \langle \gamma^2(\bm{p}) \rangle
 = \int_0^{2\pi}\frac{d\varphi_{\bm{p}}}{2\pi}\,\gamma^2(\bm{p})
 = 1.
\end{align}
The TR symmetry breaking in this state is due to the finite center-of-mass
momentum $q_x$ of the $d$-wave Cooper pairs.

Nagato and Nagai discussed the phase transition from the N state to the $q_x
= 0$ state in the film system of Fig.\ \ref{fig:Dfilm} with specular
surfaces \cite{Nag95}. At the N-SC phase boundary, $\Delta(y)$ was shown to
take the form
\begin{align}
 \Delta(y) \propto \cos(q_yy), \ \ q_y = \pi/D.
\end{align}
The spatial variation of $\Delta(y)$ originates from the pair breaking caused by
quasiparticle scattering at the film surfaces. As a consequence of the
surface pair breaking, the film system has a critical thickness at which the
N-SC phase transition occurs \cite{Nag95}. Vorontsov found that the
finite $q_x$ state has a smaller critical thickness than that of the $q_x = 0$ state
\cite{Vor09}. This means that the N state instability first occurs
for the $q_x \neq 0$ state.

In previous theories \cite{Vor09,Nag95}, the Fermi surface is assumed to
be isotropic (cylindrical). We generalize these theories to a film system
with an anisotropic Fermi surface. To discuss the effect of the Fermi surface
shape, we employ a square-lattice tight-binding model, which gives the
dispersion~\cite{Shi99,Miy98,Tan12}
\Equation{eq:2Depsilon}
{
\xi_\vp = - 2 t (\cos p_x + \cos p_y ) - \mu,
}
where $\mu$ is the chemical potential. In this model, cylindrical and square
Fermi surfaces are obtained in the limits of $\mu \rightarrow -4t$ and $\mu
\rightarrow 0$, respectively. By controlling $\mu$, we can gradually change
the Fermi surface shape from cylindrical to square (Fig.\ \ref{fig:Fermi}).

The N-SC phase boundary is determined from the gap equation
\begin{align}
 \Delta(\bm{R})\ln\frac{T}{T_c}
 &= \pi T\sum_{\epsilon_n} \Bigg\langle
 \frac{\rho_0(\varphi_{\bm{p}})\gamma(\bm{p})}
 {\langle\rho_0(\varphi_{\bm{p}})\gamma^2(\bm{p})\rangle} \notag\\
 &\times
 \left[f(\bm{R}, \bm{p}, \epsilon_n)
 - \frac{\gamma(\bm{p})\Delta(\bm{R})}{|\epsilon_n|}\right]
 \Bigg\rangle
 \label{eq:Gapeq}
\end{align}
in which the pair amplitude $f(\vR,\vp,\epsilon_{n})$ obeys the linearized
Eilenberger equation \cite{Eil68,Lar69,SerenePhysRep}
\Equation{eq:Eieq} {
  \Biggl [\epsilon_{n}
  +\frac{1}{2} {\vv}(\varphi_{\vp}) \cdot{\nabla_{\vR}}
  \Biggr ] f(\vR,\vp,\epsilon_{n})
  =
  {\rm sgn} (\epsilon_{n}) \gamma(\bm{p})\Delta(\vR).
  }
Here, $T_c$ is the transition temperature in the bulk state, $\epsilon_{n} =
\pi T (2n+1)$ is the Matsubara frequency, and $\bm{v}(\varphi_{\bm{p}})$ is the
Fermi velocity. Moreover, $\rho_0(\varphi_{\bm{p}})$ is the angle-dependent
density of states at the Fermi level defined through the replacement
\begin{align}
 \sum_{\bm{p}}(\cdots) \rightarrow
 \left<\rho_0(\varphi_{\bm{p}})
 \int d\xi_{\bm{p}} (\cdots)\right>.
\end{align}

For the self-consistent solution $\Delta(\bm{R}) \propto
\cos(q_yy)e^{iq_xx}$, the linearized gap equation is reduced to
\Equation{eq:Tceq}
{
     \ln \frac{T}{\Tc}
       =
       2\pi T
       \sum_{\epsilon_{n} > 0}  \Biggl <
       \frac{\lambda(\varphi_{\bm{p}})}
       {\bigl < \lambda(\varphi_{\bm{p}}) \bigr >}
       \operatorname{Re}
       \Biggl ( \frac{1}{\epsilon_{n} + i \eta_{\vq}}
       -\frac{1}{\epsilon_{n}}
       \Biggr ) \Biggr > ,
       }
where
\begin{align}
 \eta_{\vq}
 = \frac{1}{2}{\vv}(\varphi_{\vp}) \cdot \vq
 = \frac{1}{2}[v_x(\varphi_{\vp})q_x + v_y(\varphi_{\vp})q_y]
 \label{eq:etadef}
\end{align}
and $\lambda(\varphi_{\bm{p}}) = \rho_0(\varphi_{\bm{p}})\gamma^2(\bm{p})$.
Equation (\ref{eq:Tceq}) determines the critical thickness as a function of
$(T, q_x, \mu)$. The $\mu$ dependence comes from $\bm{v}(\varphi_{\bm{p}})$
and $\rho_0(\varphi_{\bm{p}})$. An optimum value of $q_x$ is determined
such that the critical thickness is minimized at a given $(T, \mu)$, leading to a
$D(T)$ phase boundary for a given $\mu$.

\begin{figure}[h]
\includegraphics[width=9cm]{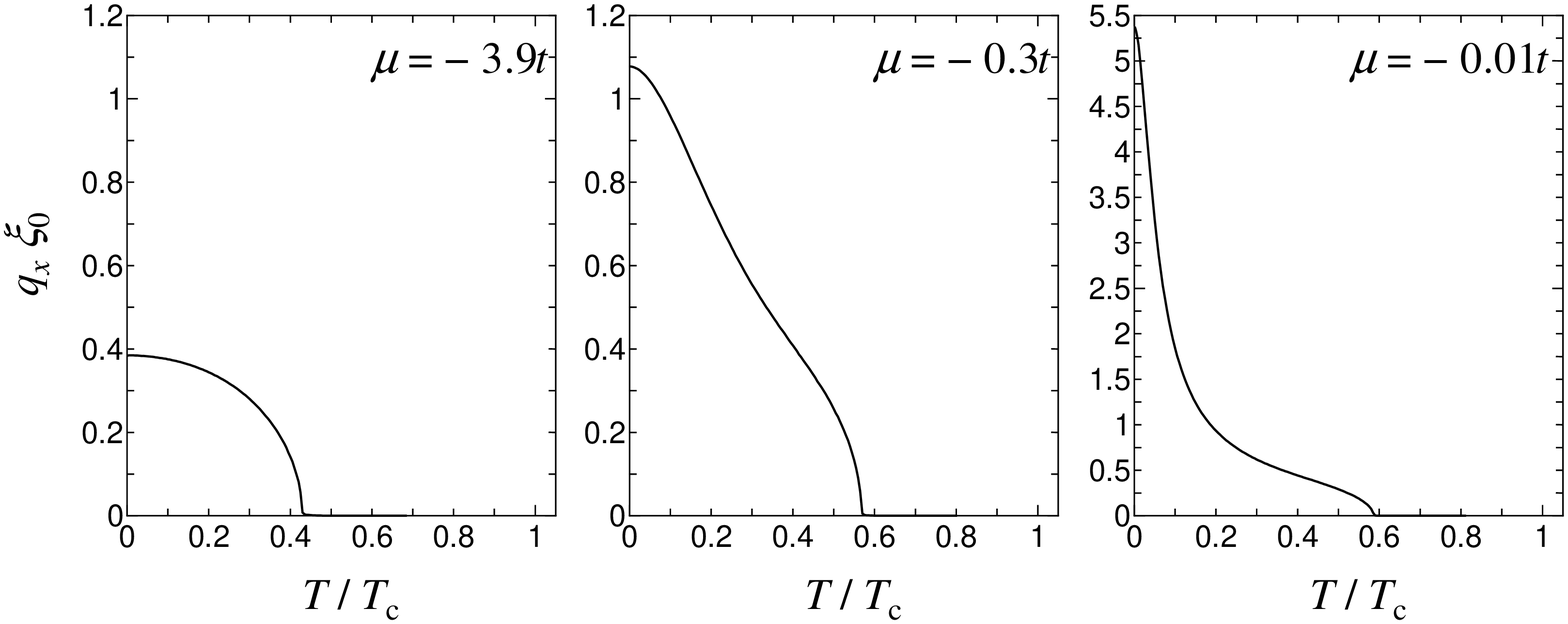}
\caption{\label{fig2} Temperature dependence of the optimum $q_x$ for $\mu
=$ $-3.9t$, $-0.3t$, and $-0.01t$.}
\end{figure}
\begin{figure}[h]
\includegraphics[width=7cm]{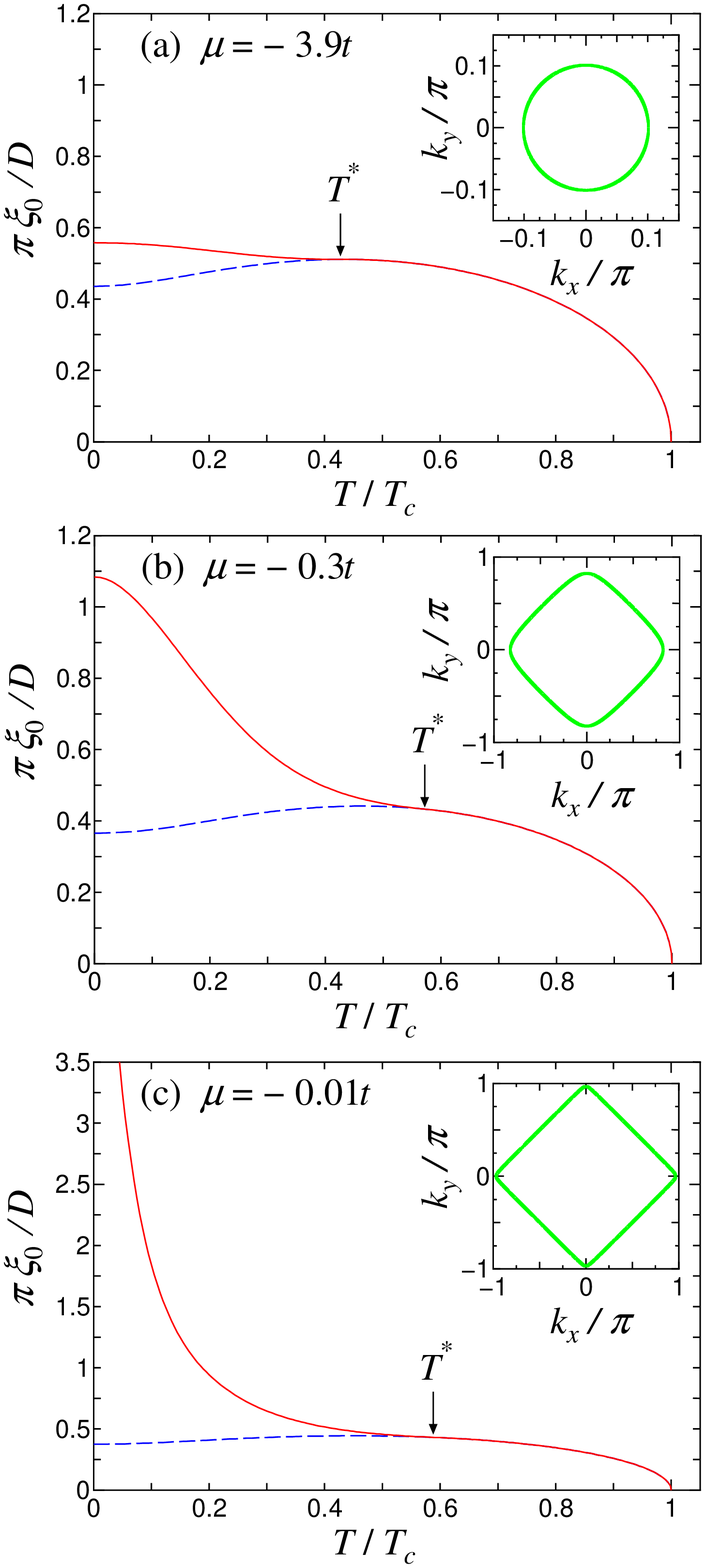}
\caption{\label{fig3} (Color online) Phase diagram for (a) $\mu=-3.9t$, (b)
$\mu=-0.3t$, and (c) $\mu=-0.01t$. The red solid (blue dashed) curve is the
phase boundary between the N state and the $q_{x} \ne 0$ ($q_{x} = 0$)
state. The arrows indicate the tricritical temperature $T^* \approx$ (a)
$0.43T_c$, (b) $0.57T_c$, and (c) $0.59T_c$.  In the inset, we depict the
Fermi surface corresponding to the $\mu$ value.}
\end{figure}

In Fig.\ \ref{fig2}, we plot the optimum $q_x$ for $\mu =$ $-3.9t$, $-0.3t$,
and $-0.01t$ as a function of $T/T_c$. The corresponding phase boundary
lines are shown in Fig.\ \ref{fig3}. The vertical axis in Fig.\ \ref{fig3}
is $\pi\xi_{0}/D = q_y\xi_0$ with the coherence length defined by
\begin{align}
 \xi_{0} = \frac{\sqrt { \bigl <
 v_{x}^{2} (\varphi_{\vp}) + v_{y}^{2} (\varphi_{\vp}) \bigr> }}{2\pi\Tc}.
\end{align}
For $\mu = -3.9t$, the Fermi surface shape is almost cylindrical, and
the phase diagram [Fig.\ \ref{fig3}(a)] quantitatively agrees with
that in Ref.\ \cite{Vor09}.  As the Fermi surface approaches the square shape,
the phase boundary between the N state and the $q_x \neq 0$ state (red solid
lines) is remarkably enhanced, and simultaneously, the tricritical
temperature $T^*$ is shifted higher.  In the case of $\mu =
-0.01t$, the phase boundary at $T=0$ is shifted to $\pi\xi_{0}/D \approx
5.37$.

\begin{figure}[h]
\includegraphics[width=8cm]{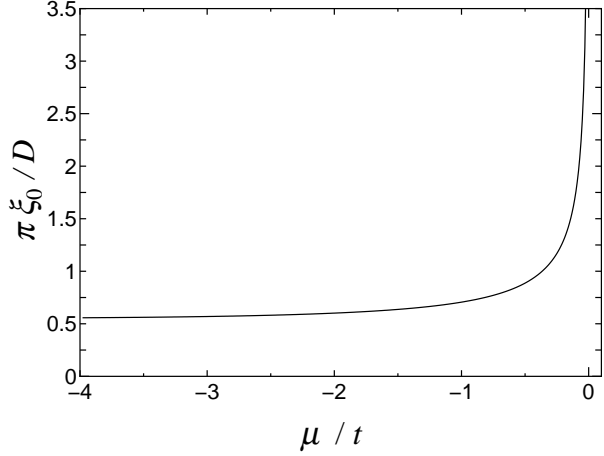}
\caption{\label{fig4} Critical value of $\pi\xi_{0}/D$ at $T=0$ as a
function of $\mu/t$.}
\end{figure}

\begin{figure}
\includegraphics[width=9cm]{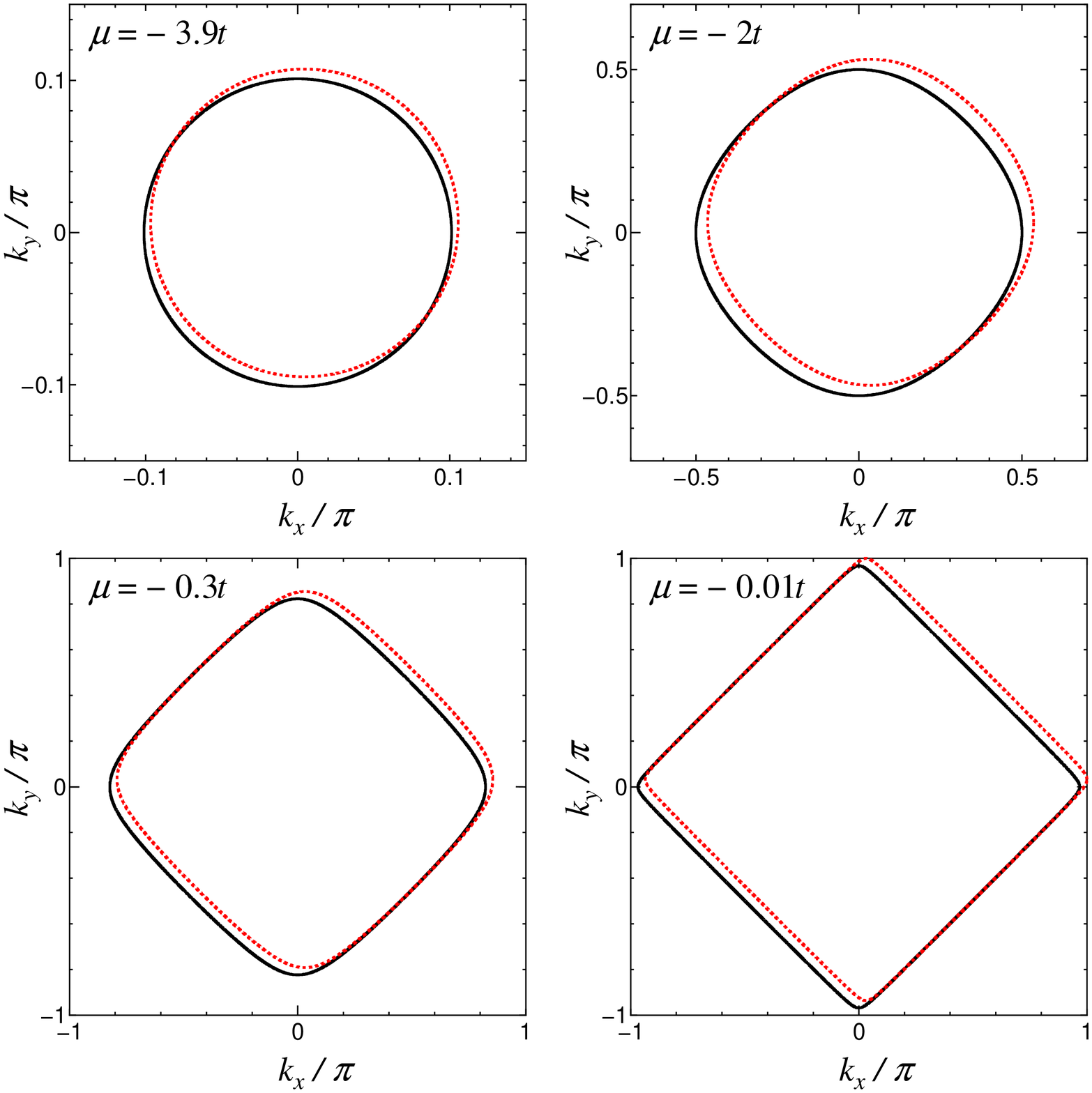}
\caption{\label{fig:Fermi} (Color online) Fermi surface nesting for
$\mu=-3.9t, -2t, -0.3t$, and $-0.01t$.  The red dotted curves show the Fermi
surface shifted by $\vq = (q_x, q_y)$.
}
\end{figure}

At $T=0$, the critical thickness is determined by
\Equation{eq:Tc0eq} {
     \begin{split}
      0
      \, = \, &
        \,\,
      \Biggl <
      \frac{\lambda(\varphi_{\bm{p}})}
      {\bigl < \lambda(\varphi_{\bm{p}}) \bigr >}
      \ln \Biggl | \frac{2 \e^{\gamma} \eta_{\vq}}{\pi \Tc} \Biggr |
      \Biggr > ,
     \end{split}
     }
where $\gamma = 0.57721\cdots$ is Euler's constant.  In Fig.\ \ref{fig4}, we
plot the critical value of $\pi\xi_{0}/D$ as a function of $\mu/t$.  As
$\mu/t$ approaches zero (the square-Fermi-surface limit), the critical
inverse thickness increases steeply and diverges in the limit $\mu/t \to 0$.


The reason for phase boundary enhancement due to the anisotropic Fermi
surface can be understood as follows. In the film system under
consideration, the gap function is suppressed by surface scattering.  In
Eq.\ \eqref{eq:Tceq}, the surface pair-breaking effect is described through
$\eta_{\bm{q}}$ with $q_x = 0$, i.e.,
\begin{align}
 \eta_{q_y} = v_y(\varphi_{\bm{p}})q_y/2 = v_y(\varphi_{\bm{p}}) \pi/2D.
 \label{eq:etaydef}
\end{align}
The finite modulation $q_x$ adds the term $\eta_{q_x} =
v_x(\varphi_{\bm{p}})q_x/2$ to $\eta_{\bm{q}}$. This additional term can
cause $\eta_{\bm{q}}$ to equal zero on some portions of the Fermi surface. The
condensation energy lost by the surface pair breaking can thus be
compensated by inducing modulation along the surface. This accounts for the stabilization of the finite $q_x$ state in the film. Note that the
condition $\eta_{\bm{q}} = 0$ on the Fermi surface is equivalent to the
nesting condition $\xi_{\bm{p}} - \xi_{\bm{p} - \bm{q}} = 0$ of the Fermi
surface.  Thus, good Fermi surface nesting helps stabilize the
finite $q_x$ state. In Fig.\ \ref{fig:Fermi}, we show the Fermi surfaces for
several $\mu$ values. The nesting condition can be satisfied in a finite
area for $\mu=-0.01t$, whereas it can only be satisfied on lines for $\mu=-3.9t$ as in the case
of the isotropic (cylindrical) Fermi surface.  As a result, the nearly
square Fermi surface ($\mu = -0.01t$) causes substantial enhancement of the
phase boundary between the N state and the $q_x \neq 0$ state.

A similar Fermi surface effect has been discussed in the context of FFLO
instability \cite{Shi94,Shi99,Miy14}. In that case, the surface pair-breaking term
$\eta_{q_y}$ in $\eta_{\bm{q}}$ is replaced by the Zeeman coupling to
an external magnetic field, and the critical inverse thickness corresponds to
the upper critical field $H_{c2}$. As shown with the FFLO
problem for Q1D systems \cite{Buz83,Suz83,Mac84}, the $H_{c2}(T)$ curve has a
positive curvature in the case of perfect nesting. The corresponding
behavior is found in Fig.\ \ref{fig2}(c) ($\mu = -0.01t$) in the upturn of
the red solid line below $T^*$. When $\mu = -0.3t$, similar upturn behavior
is found; however, in this case, the curvature becomes negative at low
temperatures. The low-temperature difference occurs because the nesting
condition for $\mu = -0.3t$ is not ``touching on surfaces'' but ``crossing
on lines.'' At high temperatures, the thermal energy $k_BT$ makes the
difference between the two conditions indistinguishable \cite{Miy14}, and the phase boundary line exhibits an upturn similar to the case of $\mu =
-0.01t$.

In conclusion, we have discussed the effect of the Fermi surface shape on the
N-SC phase transition in unconventional SC films with pair-breaking
surfaces. We have demonstrated the Fermi surface effect using the phase transition to the TR symmetry breaking state predicted for
a $d$-wave SC film \cite{Vor09} as an example, and we showed that the critical thickness is
substantially reduced near half filling, where the Fermi surface is almost
square shaped. This result can be interpreted as a consequence of Fermi
surface nesting. An analogous Fermi surface effect has been discussed for
the FFLO stability problem \cite{Shi94,Shi99,Miy14}.  In this case, Fermi surface
nesting causes an enhancement of the upper critical field $H_{c2}$. In both
cases, the spatial modulation of the gap function is induced to avoid strong
pair breaking. The pair breaking is caused by surface scattering in the film
case or by an external magnetic field in the FFLO case.
In general, Fermi surface nesting allows a much greater reduction in the pair-breaking effect.
The stabilization
scenario for nonuniform superconductivity by the Fermi surface nesting can
be applied quite generally to superconductors with strong pair breaking.

We would like to thank H. Shimahara, Y. Nagato, S.-I. Suzuki, and K. Nagai
for their helpful discussions and comments.  This work was supported in part
by the ``Topological Quantum Phenomena'' (No. 22103003) Grant-in-Aid for
Scientific Research on Innovative Areas from the Ministry of Education,
Culture, Sports, Science and Technology (MEXT) of Japan.

\end{document}